\documentclass[12pt]{article}
\usepackage{amssymb}
\begin{document}
\baselineskip=18pt \pagenumbering{arabic}
\parskip1.5em
\thispagestyle{empty} \vskip2.5em
\begin{center}
{\Large{\bf An Algorithm for the Microscopic Evaluation of the Coefficients of the Seiberg-Witten
Prepotential}}\\\vskip1.5em R. Flume$^{a}$\hspace{3em}R. Poghossian$^{b}$
\\ \vskip3em
$^{a}${\sl Physikalisches Institut der Universit\"at Bonn}\\
{\sl Nu{\ss}allee 12, D--53115 Bonn, Germany}\\
\vskip2em $^{b}$ {\sl Yerevan Physics
Institute}\\
{\sl Alikhanian Br. st. 2, 375036 Yerevan, Armenia}
\end{center}
\begin{abstract}
\noindent A procedure, allowing to calculate the coefficients of
the SW prepotential in the framework of the instanton calculus is
presented. As a demonstration explicit calculations for $2$, $3$
and $4-$ instanton contributions are carried out.
\end{abstract}

\begin{center}
To the memory of Lochlainn O'Raifeartaigh
\end{center}

\vspace{2 cm}
{\small{
e-mail: flume@th.physik.uni-bonn.de\\
\hspace*{1.85 cm} poghos@moon.yerphi.am}}
\vfill\eject\setcounter{page}{1}

\setcounter{equation}{0}
\section{Introduction}
\renewcommand{\theequation}{1.\arabic{equation}}
One of the challenges put forward by the Seiberg-Witten proposal
\cite{sw1} for an exact expression of the prepotential in $N=2$
extended $d=4$ Super Yang-Mills theory is its verification through
microscopic instanton calculations. Dorey, Khoze and Mattis made
the first steps into that direction determining the 1 and
2-instanton contribution to the prepotential \cite{dorey1}. (The
1-instanton contribution was already calculated before in
references \cite{finnel} and \cite{ito}.) These authors also
derived a representation of the integration measure of instanton
moduli in general \cite{dorey3}. After this work further progress
was hampered by the fact that beyond instanton number 3 the ADHM
(Atiyah, Drinfeld, Hitchin, Manin \cite{adhm}) moduli space is not
known explicitly. In \cite{FPS1}, \cite{FPS2} we derived a
standard algebraic-geometric form, to be reproduced below,
representing the expansion coefficients of the prepotential as
integral of the exponential of an equivariantly exact mixed
differential form. This can be reduced  in the $k$-instanton
sector to an integral of a (4k-3)-fold wedge product of a closed
differential 2-form, which is the formal representative of the
Euler class of the moduli space viewed as U(1) principal bundle.
So we were led to an algebraic-geometric interpretation of the
coefficients. But we failed in our main objective, which was to
calculate via localization technique concrete numbers. The reason
for our failure was clarified and cured in a recent work by
Hollowood \cite{H}, where it is argued that one should, to start
with, resolve the short distance singularities of the moduli
space\footnote{ R.F. thanks Francesco Fucito for a discussion of
this point.}.

An elegant method to achieve the resolution consists in deforming
the instanton configurations into a non-commutative domain
\cite{NSch}. The parameter measuring the non-commutativity, a new
length scale, acts as an ultraviolet regulator, while the
integrals in question are - as formally shown below- insensitive
to the deformation. The integrals can now be localized at the
surface of critical points of the abelean vector field which goes
with the remaining torus symmetry after spontaneous breakdown of
the non-abelean gauge symmetry of the N=2 Yang-Mills
theory\footnote{Localization without regularization renders a
vanishing residuum at the corresponding critical surface.}.

The critical surface happens to coincide with the the manifold of
non-commutative U(1) instantons (imbedded as submanifold into the
ADHM moduli space of the non-abelean gauge group). The remaining
integral along the critical surface has according to the recipes
of the equivariant localization technique \cite{B}, \cite{BGV} the
appearance of a sum of characteristic classes of the unitary
bundle connected with the ADHM construction. Their evaluation is
for general instanton number still a highly cumbersome task.

A conceivable way to overcome the difficulties relies on the
deformation invariance of the characteristic classes. So one may
follow backwards a gradient flow along a Gl(k) orbit \cite{DK}
(chapter 6), \cite{Nak}, which relates the relevant matrices of
the ADHM construction to diagonal matrices. We will not follow
this approach here since we were reminded meanwhile by Nekrasov
\cite{N} of a technically simpler method\footnote{We still believe
that it is worthwhile to pursue attempts either to evaluate the
integrals without regularization or to handle the U(1)-instanton
manifolds. It may lead to new mathematical insights.}. The idea,
already used some time ago in a macroscopical context
\cite{MNSch1}, \cite{MNSch2}, is to modify the vector field
underlying the localization method by adding pieces corresponding
to space rotations. The critical set of the modified vector field
consists of discrete points and the evaluation of the integrals is
reduced to a manageable task: to find the eigenvalues of the
vector field action on the tangent space of the respective
critical point. The purpose of this paper is to show that the
program can be implemented on the microscopic level following the
conceptual lines of our previous papers \cite{FPS1}, \cite{FPS2}.
Similar results have been announced in \cite{N}. We collect in the
following section 2 some material concerning the instanton
calculus. Section 3 is devoted to a description of the algorithm
allowing to determine (in principle) the contributions from
arbitrary instanton sectors. We will work out for the purpose of
illustration the 2 ,3 and 4-instanton coefficient of the pure N=2
vector theory (i.e. without matter multipletts) with SU(2) as
underlying gauge group in section 4. We end this section
presenting a formula for the determinant of the deformed vector
field action on the neighborhood of an arbitrary critical point in
the general case of $SU(N)$ as gauge group.

\setcounter{equation}{0}
\section{Instanton calculus}
\renewcommand{\theequation}{2.\arabic{equation}}
We quote the ADHM data for the construction of $SU(2)$ instantons
in a form which is generalizable to $SU(N)$, $N\geq 2$. (The
restriction to $SU(2)$ is a matter of convenience; all our results
can easily be extended to $SU(N)$). These data consist of complex
matrices, two $k\times k$ matrices $B_1$, $B_2$, a $k\times 2$
matrix $I$ and a $2\times k$ matrix $J$, fulfilling a certain rank
condition (cf. \cite{adhm}, \cite{DKM'}) and obeying the relations
\begin{eqnarray}
\label{complexADHM}
\left[ B_1, B_2\right]+IJ=0; \\
\label{realADHM} \left[ B_1, B_1^{\dagger}\right]+\left[ B_2, B_2^{\dagger}\right]+II^{\dagger}-J^{\dagger}J=0,
\end{eqnarray}
where $"^{\dagger}"$ denotes hermitean conjugation. The data are redundant in the sense that sets of matrices
related by $U(k)$ transformations,
\begin{eqnarray}
{\cal A}\equiv \left(B_1, B_2, I, J\right)\leftrightarrow {\cal A}^{\prime}, \nonumber \\
B_i^{\prime}=gB_ig^{-1}, \, \, \, I^{\prime}=gI, \, \, \, J^{\prime}=Jg^{-1}; \nonumber \\
i=1,2; \, \, \, \, g\in U(k)
\end{eqnarray}
give rise to the same self dual Yang-Mills configuration of
topological charge $k$. Factoring out the redundancy from the
above data (e.g. by imposing gauge fixing conditions) one is lead
to the instanton moduli space as the set of gauge equivalence
classes of solutions of Eq.'s (\ref{complexADHM}),
(\ref{realADHM}). This space is not smooth. It contains boundary
points where the previously mentioned rank condition is violated
(which corresponds to Yang-Mills configurations with topological
charge density concentrated in part at single points). A
resolution of singularities is achieved \cite{Nak} through a
modification of Eq. (\ref{realADHM}):
\begin{equation}
\label{realADHMdeformed} \left[ B_1, B_1^{\dagger}\right]+\left[ B_2,
B_2^{\dagger}\right]+II^{\dagger}-J^{\dagger}J=\zeta \, {\bf 1}_{k \times k},
\end{equation}
where $\zeta$ denotes a positive real number. Eq.
(\ref{realADHMdeformed}) is in fact the starting point for the
construction of solutions of the Yang-Mills selfduality  equation
on a non-commutative space-time \cite{NSch} with $\zeta$ setting
the scale of the non-commutativity. The appearance of the
non-commutative space will be of no relevance for our purposes.
Going from Eq. (\ref{realADHM}) to Eq. (\ref{realADHMdeformed}) we
will content ourselves with an argument, to be given below, that
the numbers we want to calculate do not depend on $\zeta$.

Dealing with a supersymmetric theory in quasi-classical
approximation one has to take into account fermionic degrees of
freedom which are here the Weyl zero modes of positive chirality
in the selfdual Yang-Mills background. It turns out, \cite{FPS1},
\cite{FPS2}, that a neat realization of the fermions is supplied
on the level of moduli through their identification with a basis
of the cotangent bundle of these moduli. To adapt the cotangent
space in a $U(k)$-invariant way we introduce some notation. Let
${\bf L}$ be a selfadjoint operator acting on the space of
anti-hermitean $k \times k$ matrices $M$ by
\begin{eqnarray}
{\bf L}\cdot M=\left\{ II^{\dagger}+J^{\dagger}J,M
\right\}+\sum_{l=1,2}\left( \left[B_l, \left[B_l^{\dagger}, M
\right]\right]+\left[B_l^{\dagger}, \left[B_l, M\right]
\right]\right).
\end{eqnarray}
It can be shown that ${\bf L}$ is invertible (in fact positive) as
long as the above mentioned rank condition is satisfied (under the
supposition of Eq.'s (\ref{complexADHM}), (\ref{realADHM}). The
invertibility is guaranteed per se if one relies on Eq.'s
(\ref{complexADHM}), (\ref{realADHMdeformed}) ). Let ${\bf X}$ be
the matrix valued differential one-form
\begin{eqnarray}
{\bf X}=\sum_{i=1,2}\left[B_i^{\dagger},dB_i\right]+J^{\dagger}dJ-dII^{\dagger}-h.c.
\end{eqnarray}
We introduce a $U(k)$-covariantized exterior derivative on the ADHM matrices ${\cal A}\equiv (B_1, B_2, I, J)$
setting
\begin{eqnarray}
{\cal D}{\cal A}\equiv d{\cal A}+Y\cdot {\cal A},
\label{covderivative}
\end{eqnarray}
with $Y={\bf L}^{-1}{\bf X}$ and $Y\cdot {\cal A}\equiv
(\left[Y,B_1\right], \left[Y,B_2\right], YI, -JY)$. ${\cal D}{\cal
A}$ satisfies by construction (cf. \cite{FPS1}, \cite{FPS2}), the
fermionic part of the supersymmetrized ADHM conditions
(\ref{complexADHM}) and (\ref{realADHM}) (or
(\ref{realADHMdeformed}) resp.) thus giving rise to Weyl zero
modes. It is worth to mention the geometric meaning of the $U(k)$
connection field $Y$. The flat euclidean space of data ${\cal A}$
(without Eq.'s (\ref{complexADHM}), (\ref{realADHM})
((\ref{realADHMdeformed})) can be supplied with a K\"{a}hler
metric. The K\"{a}hler property is preserved by the restriction
imposed by Eq.'s (\ref{complexADHM}), (\ref{realADHM})
((\ref{realADHMdeformed})) provided one projects in the tangent
bundle of the moduli manifold onto the lifted (with respect to the
$U(k)$ action) horizontal subspace \cite{HKLR}, \cite{BF}. To come
to terms with the latter notion let us assume that the moduli
space is realized through the imposition of a gauge fixing
condition, e.g. by demanding that, say, the hermitean part of
$B_1$ is diagonal. The tangent space of the larger manifold
(without gauge fixing) may be decomposed into the subspace spanned
by the infinitesimal $U(k)$ generators, the so called "vertical"
subspace and its orthogonal\footnote{orthogonal with respect to
the induced metric.} complement, the "horizontal" subspace, the
latter being isomorphic to the tangent space of the gauge fixed
moduli manifold ${\cal M}_k$. Given a tangent vector $T$ on ${\cal
M}_k$ one finds a unique vector $V$ in the vertical subspace s.t.
$(V+T)$, called "the horizontal lift of $T$", is contained in the
horizontal space. All that extends naturally to the corresponding
cotangent bundles. The covariant derivative introduced in Eq.
(\ref{covderivative}) serves just this purpose: it lifts the
ordinary exterior derivative on ${\cal M}_k$ to the horizontal
subspace of the larger cotangent bundle. A natural metric ${\tilde
g}$ on the horizontal subspace induced from the flat one is given
by
\begin{equation}
{\tilde g}\left(d {\cal A}^{\dagger}, d {\cal A}\right)=g \left({\cal D} {\cal A}^{\dagger}, {\cal D} {\cal
A}\right),
\end{equation}
$g$ being the flat metric
\begin{equation}
g\left(d {\cal A}^{\dagger}, d {\cal A}\right)=tr \left( \, \,  \sum_{l=1,2} dB_l^{\dagger} dB_l +dI^{\dagger}
dI+dJ dJ^{\dagger}\right).
\end{equation}

One of the results of \cite{FPS1}, \cite{FPS2} is that the coefficient ${\cal F}_k$ of the ${\cal N}=2$
prepotential
\begin{equation}
{\cal F}(\Psi)=\frac{i}{2\pi} \Psi^2\log \frac{2\Psi^2}{e^3\Lambda^2} -\frac{i}{\pi}\sum_{k=1}^{\infty}{\cal
F}_k \left(\frac{\Lambda}{\Psi} \right)^{4k} \Psi^2, \label{Fexpansion}
\end{equation}
quoted here as a function of the ${\cal N}=2$ chiral vector superfield $\Psi$ can be represented as an integral
over the reduced moduli space $ {\cal M}'_k$ - that is the space from which the coordinates of the instanton
center and its fermionic counterparts have been eliminated,
\begin{equation}
{\cal F}_k \simeq \int_{{\cal M}'_k} e^{-d_x\omega}, \label{modspaceint}
\end{equation}
where $\omega$ is here the differential one-form
\begin{equation}
\omega=\Re e \left({\cal D}I \bar{v} I^{\dagger} +J^{\dagger} v
{\cal D}J \right), \label{omega}
\end{equation}
$v$ denotes the Higgs field (the scalar component of the above mentioned superfield $\Psi$) vacuum expectation
value breaking  the gauge group $SU(2)$ down to $U(1)$. We will chose $v$ to be of the form
\begin{equation}
v=\left(
\begin{array}{cc}
ia & 0 \\
0 & -ia \end{array}
\right),
\end{equation}
where $a$ is real.
 $d_x$ in Eq. (\ref{modspaceint}) stands for an
equivariant differential
\begin{equation}
d_x\equiv d+i_{x},
\end{equation}
$d$ being the ordinary exterior derivative and $i_x$ meaning contraction with the $U(1)$ vector field (denoted
below by $x$ ) going along with the infinitesimal transformation
\begin{eqnarray}
\delta B_i \sim 0; \, \, \, \, \delta I\sim Iv; \, \, \, \delta J\sim -vJ.
\label{vectorfield}
\end{eqnarray}
It is worthwhile to note, that the one-form $\omega$ is dual to
the vector field (\ref{vectorfield}) with respect to the metric
${\tilde g}$:
\begin{equation}
\omega =\Re e \,{\tilde g}\left(x^{\dagger}, d{\cal A} \right).
\label{omega'}
\end{equation}
The coefficient ${\cal F}_k$ may be deformed into
\begin{equation}
{\cal F}_k(t) \equiv \int_{{\cal M}'_k} e^{-\frac{1}{t}d_x\omega} \label{tdeformedmodspaceint}
\end{equation}
and we obtain
\begin{equation}
\label{tindependence1}
\frac{d}{dt}{\cal F}_k(t) =-\frac{1}{t^2} \int_{{\cal M}'_k}d_x \left(\omega e^{-\frac{1}{t}d_x\omega}\right) = \\
\label{tindependence2}
 =-\frac{1}{t^2} \int_{{\cal M}'_k}d \left(\omega
 e^{-\frac{1}{t}d_x\omega}\right).
\label{tdeformedmodspaceint}
\end{equation}
For the equality (\ref{tindependence1}) use has been made of the equivariance of the one-form $\omega$,
\begin{equation}
d_x^2 \omega=\left(d \circ i_x+i_x \circ d\right)\omega={\cal L}_x
\omega=0,
\end{equation}
where ${\cal L}_x$ is the Lie-derivative of the above introduced
$U(1)$ vector field. To understand the second equality in
(\ref{tindependence2}) one has to note that the the integral over
${\cal M}'_k$ has to be taken with the top form from $\exp
{-(\frac{1}{t}d_x\omega )}$ inserted, which can be reached (being
a top form ) only by $d$, not by $i_x$. The $t$-dependence of
${\cal F}_k(t)$ hinges on whether the total derivative integral in
(\ref{tindependence2}) picks up boundary terms. Boundary terms are
apparently present in the unregularized version of ${\cal M}'_k$
defined by Eq.'s (\ref{complexADHM}), (\ref{realADHM}) (see
\cite{FPS1}, \cite{FPS2}), but are not present, as has been
convincingly argued by Hollowood \cite{H}, after
$\zeta$-regularization\footnote{One may worry that the
non-compactness of ${\cal M}'_k$ may create problems. This is not
the case . Estimates involving the dilute gas approximation for
instantons show that the integrands decay fast enough into the
non-compact directions to avoid trouble.}. ${\cal F}_k$ does
nevertheless not depend on $\zeta$. Indeed, the integral
(\ref{tdeformedmodspaceint}) over manifolds ${\cal M}'_k$
associated to different parameters $\zeta$ are related by a simple
rescaling which can be absorbed into a change of the parameter $t$
and so does not, according to Hollowood, alter ${\cal F}_k$.

The result
\begin{equation}
\left. \frac{d {\cal F}_k}{d t}\right|_{\zeta \neq 0}=0 \nonumber
\end{equation}
suggests a saddle point evaluation of the integral over ${\cal
M}'_k$, which should render an exact result. The saddle points are
determined by $i_x \omega$,
\begin{eqnarray}
i_x \omega= {\tilde g}\left(x^{\dagger}, x\right)=tr \left(
\sum_{l=1,2}\left[{\bf L}^{-1}\Lambda,
B_l^{\dagger}\right]\left[{\bf L}^{-1}\Lambda, B_l\right]\right.
\nonumber \\ + \left(-vI^{\dagger}-I^{\dagger}{\bf
L}^{-1}\Lambda\right)\left(I v+{\bf L}^{-1}\Lambda I\right)
\nonumber
\\+\left. \left(J^{\dagger}v+{\bf L}^{-1}
\Lambda J^{\dagger}\right)\left( -vJ-J{\bf L}^{-1}\Lambda \right)\right).
\label{bosonicpart}
\end{eqnarray}
The r.h.s. of Eq. (\ref{bosonicpart}) is a sum of squares, each of
which should vanish for itself at a saddle point. In this way we
are lead to the saddle point equalities
\begin{eqnarray}
\left[{\bf L}^{-1}\Lambda, B_l\right]=0; \nonumber \\
I v+{\bf L}^{-1}\Lambda I=0; \nonumber \\
-vJ-J{\bf L}^{-1}\Lambda =0. \label{saddlepointconditions}
\end{eqnarray}
One should note that the ${\bf L}^{-1}\Lambda$ appearing in these
equations are emerging from the $U(k)$ connections of the $U(1)$
vector field due to the horizontal lift of the latter. The
interpretation of Eq.'s (\ref{saddlepointconditions}) is obvious:
The saddle points are identical with the places at which the
lifted vector field vanishes and can be shown, \cite{H}, to
coincide with the union of direct products of two $U(1)$ instanton
configurations with topological charges $k_1$ and $k_2$ satisfying
the condition $k_1+k_2=k$. Using the general localization formula
(see \cite{BGV}, p. 216 ff.) we can reduce the $8k-4$ dimensional
integral (\ref{modspaceint}) over the moduli space ${\cal M}'_k$
to a sum of $4k_1 +4k_2-4=4k-4$ dimensional integrals over the
spaces ${\cal M}'_{k_1 , k_2}\equiv {\cal M}_{k_1}(U(1)) \times
{\cal M}_{k_2}(U(1))\backslash {\bf C}^2$ (the common center of
two $U(1)$ instantons is eliminated)
\begin{equation}
\int_{{\cal M}'_k} e^{-d_x\omega}=\sum_{\sb {k_1, k_2} \atop k_1+k_2=1} \int_{{\cal M}_{k_1 ,
k_2}}\frac{1}{\det^{1/2} \left({\cal L}_{\cal N}+{\cal R}_{\cal N}\right)}, \label{normalbundleint}
\end{equation}
where ${\cal L}_{\cal N}$ is the action of the vector field on the
subspaces of horizontal spaces orthogonal to ${\cal M}_{k_1 ,
k_2}$ (these subspaces constitute the so called normal bundle) and
${\cal R}_{\cal N}$, a two form along ${\cal M}_{k_1 , k_2}$ and a
linear operator acting on these orthogonal subspaces, is the
curvature of the normal bundle. Eq. (\ref{normalbundleint}) shows
that each SW coefficient ${\cal F}_k$ is equal to a certain
polynomial of the characteristic classes of the normal bundle. The
formula (\ref{normalbundleint}) providing a drastic simplification
of the initial problem, still does not allow us to perform
computations beyond $2$-instantons\footnote{The $2$-instanton
calculation goes smoothly and gives the desired result.}.

\setcounter{equation}{0}
\section{The modified vector field }
\renewcommand{\theequation}{3.\arabic{equation}}
We have seen in section 2, that the main building block of the
superinstanton action is the vector field $x$ (see
(\ref{modspaceint}), (\ref{omega}), (\ref{omega'})). The
difficulty we have encountered was a too large zero set of this
vector field. We describe a natural deformation of this vector
field whose zero set is only a discrete finite set. Such a
vectorfield has been for good use in mathematics (cf. \cite{Nak},
and references therein\footnote{We became aware of the relevance
of Nakajima's work through \cite{N}.}). We will heavily rely on
some of the results of Nakajima in the following. Consider two
independent rotations of space-time, first  on the $x_1$, $x_2$
plane with the rotation angle $\epsilon_1$ and the second one on
the plane $x_3$, $x_4$ with rotation angle $\epsilon_2$. It is
convenient to introduce complex coordinates $z_1=x_1+ix_2$,
$z_2=x_3+ix_4$ in (euclidean) space-time. The group element
specified by the parameters $(t_1,t_2)$ ($t_l\equiv \exp
i\epsilon_l$, $l=1,2$) acts on $(z_1,z_2)$ as
$(z_1,z_2)\rightarrow (t_1 z_1, t_2 z_2)$. The respective action
on the ADHM data is given by:
\begin{equation}
B_l\rightarrow t_l B_l ; \, \, \, \, I\rightarrow I; \, \, \, \, J\rightarrow t_1 t_2 J.
\label{torusaction}
\end{equation}
The unbroken $U(1)$ subgroup of the gauge group acts as:
\begin{equation}
B_l\rightarrow B_l ; \, \, \, \, I\rightarrow It_v; \, \, \, \, J\rightarrow t_v^{-1} J,
\label{u1action}
\end{equation}
where $t_v=\exp ia \sigma_3$ (our vector field $x$ is just the
generator of the transformations (\ref{u1action}) ). It is evident
that the transformations (\ref{u1action}), (\ref{torusaction}) act
properly also on the moduli space ${\cal M}_k$, because they
commute with $U(k)$. Let us combine the transformations
(\ref{u1action}), (\ref{torusaction})
\begin{equation}
B_l\rightarrow t_l B_l ; \, \, \, \, I\rightarrow It_v; \, \, \,
\, J\rightarrow t_1 t_2 t_v^{-1}J
\label{combinedaction}
\end{equation}
and consider its generating vector field - to be denoted ${\tilde
x}$ - in the construction of a modified superinstanton action as
the main building block substituting the vector field $x$. The
components of the vector field ${\tilde x}$ are determined through
the infinitesimal form of the transformations
(\ref{combinedaction})
\begin{equation}
\delta B_l\sim \epsilon_l B_l ; \, \, \, \, \delta I\sim Iv; \, \,
\, \, \delta J\sim -\epsilon vJ, \label{tildex}
\end{equation}
where $\epsilon=\epsilon_1+\epsilon_2$. To find the deformed
one-form ${\tilde \omega}$ we substitute in Eq. (\ref{omega'}) $x$
by ${\tilde x}$. As a result we get
\begin{equation}
{\tilde \omega} =\Re e \, tr\left(-i\sum_{l=1,2}\epsilon_l {\cal
D}B_l B_l^{\dagger}-{\cal D}I v
I^{\dagger}+J^{\dagger}\left(v-i\epsilon \right){\cal D}J\right).
\end{equation}
For the bosonic part of the deformed superinstanton action $d_{{\tilde x}} {\tilde \omega}$ we obtain
\begin{eqnarray}
i_{{\tilde x}} {\tilde \omega}= {\tilde g}\left({\tilde
x}^{\dagger}, {\tilde x}\right)=tr \left(
\sum_{l=1,2}\left(-i\epsilon_l B_l^{\dagger}+\left[{\bf
L}^{-1}{\tilde \Lambda},
B_l^{\dagger}\right]\right)\left(i\epsilon_l B_l+\left[{\bf
L}^{-1}{\tilde \Lambda}, B_l\right]\right)\right. \nonumber \\ +
\left(-vI^{\dagger}-I^{\dagger}{\bf L}^{-1}{\tilde
\Lambda}\right)\left(I v+{\bf L}^{-1}{\tilde \Lambda} I\right)
\nonumber \\+\left. \left(J^{\dagger}\left(v-i\epsilon
\right)+{\bf L}^{-1}{\tilde
\Lambda}J^{\dagger}\right)\left(\left(-v+i\epsilon\right)J-J{\bf
L}^{-1}{\tilde \Lambda} \right)\right),
\end{eqnarray}
where
\begin{equation}
{\tilde \Lambda}=i_{\tilde x} {\bf X}=\sum_{l=1,2} i\epsilon_l
\left[B_l^{\dagger},
B_l\right]+J^{\dagger}\left(-v+i\epsilon\right)J-IvI^{\dagger}-h.c.
\, \, .
\end{equation}
The deformed super-instanton action has an explicit dependence on
the instanton center. Hence we are not allowed to restrict the
moduli space of instantons imposing the condition $tr B_1=tr
B_2=0$ which means that the center of instantons has zero
coordinates.  We define the deformed version of the Eq.
(\ref{modspaceint}) as
\begin{equation}
Z_k \left(a, \epsilon_1, \epsilon_2 \right)\equiv\int_{{\cal M}_k} e^{-d_{\tilde x}{\tilde \omega}},
\label{deformedmodspaceint}
\end{equation}
where integration is over the entire moduli space ${\cal M}_k$.
The evaluation of this integral by means of localization technic
is much simpler, because on the right hand side of the
localization formula (\ref{normalbundleint}) we will have a finite
sum over the zero locus set of vector field ${\tilde x}$ and
therefore the curvature term of the normal bundle ${\cal R}_{\cal
N}$ is absent. A zero locus of the vector field ${\tilde x}$
(equivalently a fixed point of the combined action
(\ref{combinedaction})) on the moduli space) is defined by the
conditions:
\begin{equation}
t_l B_l=g^{-1}B_lg ; \, \, \, \, It_v=g^{-1}I; \, \, \, \, t_1 t_2
t_v^{-1}J=J g. \label{fixedpointconditions}
\end{equation}
Using the freedom of $U(k)$ transformations we will assume that
the group element $g$ out of $U(k)$ is diagonal
\begin{equation}
g=\left( \begin{array}{ccc} e^{i\Phi_1} & \cdots & 0 \\
 & \cdots &  \\
0 & \cdots & e^{i\Phi_k}
\end{array}
\right). \label{diagonaluk}
\end{equation}
The fixed point condition (\ref{fixedpointconditions}) implies:
\begin{eqnarray}
\delta B_{l,ij}\equiv \left(\Phi_{ij}+\epsilon_l\right)B_{l,ij}=0; \nonumber \\
\delta I_{i \lambda}\equiv \left(\Phi_i+a_{\lambda}\right)I_{i\lambda}=0; \nonumber \\
\delta J_{\lambda i}\equiv \left(-\Phi_i-a_{\lambda}+\epsilon \right)J_{\lambda i}=0; \nonumber \\
i,j=1,2, \cdots, k; \, \, \, \, l, \lambda=1,2; \, \, \, \,
\Phi_{ij}\equiv \Phi_i-\Phi_j. \label{fixedpointconditions'}
\end{eqnarray}
The geometrical meaning of these equations (as before for $x$) is
the criticality of the horizontally lifted vector field ${\tilde
x}'$ whose action  in the neighborhood  of a critical point is
given by equation
\begin{eqnarray}
\delta_{{\tilde x}'} B_{l,ij}\equiv \left(\Phi_{ij}+\epsilon_l\right)B_{l,ij} \, \, ; \nonumber \\
\delta_{{\tilde x}'} I_{i \lambda}\equiv \left(\Phi_i+a_{\lambda}\right)I_{i\lambda} \, \, ; \nonumber \\
\delta_{{\tilde x}'} J_{\lambda i}\equiv
\left(-\Phi_i-a_{\lambda}+\epsilon \right)J_{\lambda i} \, \, .
\label{xtildeaction}
\end{eqnarray}
The classification of fixed points of the torus action
(\ref{torusaction}) in the case of $U(1)$ (non-commutative)
instantons can be found in \cite{Nak} chapter 5.2. Fortunately, to
adapt this to our case of $SU(2)$ (or $SU(N)$ in general)
instantons and the combined action (\ref{combinedaction}) only
some minor modifications are needed.

There is one-to-one correspondence between the fixed points of
${\tilde x}'$ and ordered pairs of Young diagrams $(Y_1,Y_2)$ with
$k_1$ and $k_2$ boxes resp. , s.t. $k_1+k_2=k$. Denote the number
of boxes in the rows of the Young diagram $Y_l$ by $\nu_{l,1}\geq
\nu_{l,2} \geq \cdots \geq 0 $ and the number of boxes in
columns\footnote{Our convention is to numerate rows beginning from
the top to the bottom and columns from left to right.} by
$\nu'_{l_,1}\geq \nu'_{l,2} \geq \cdots \geq 0 $. Then we
distribute the phases $\Phi_1, \Phi_2, \cdots, \Phi_{k_1}$ in the
Young diagrams $Y_1$ and $\Phi_{k_1+1}, \Phi_{k_1+2}, \cdots,
\Phi_{k}$ in $Y_2$ subsequently filling row after row beginning
from the upper left corner boxes. To the phase $\Phi_m$
distributed in a box of $Y_l$ in the $i$-th row and $j$-th column
we assign the value
\begin{equation}
\Phi_m=-a_l-(i-1)\epsilon_2-(j-1)\epsilon_1, \label{Phivalues}
\end{equation}
where $a_1=-a_2=a$. The matrix element $B_{1,mn}$ ($B_{2,mn}$) is
nonzero if and only if $\Phi_m$ and $\Phi_n$ are neighbors in a
row (column). The only nonzero matrix elements of $I$ are
$I_{1,1}$ and $I_{k_1+1,2}$. $J$ vanishes identically. These are
some special points belonging to the manifold ${\cal M}_{k_1,k_2}$
introduced earlier. It is easy to see that the conditions
(\ref{fixedpointconditions'}) are fulfilled. Of course, to
determine the actual values of nonzero matrix elements one should
impose the ADHM equations (\ref{complexADHM}),
(\ref{realADHMdeformed}) but this is a more delicate problem.
Below we will solve this problem for instanton charges $k=2,3,4$.

The next step is the construction of tangent spaces passing
through the fixed points and to calculate the determinants of
${\tilde x}'$-action on this spaces. To do this, one needs to find
all solutions of the linearized ADHM equations (the so called
fermionic ADHM equations)
\begin{eqnarray}
\left[ \delta B_1, B_2\right]+\left[B_1, \delta B_2\right]+\delta IJ+I\delta J=0; \nonumber \\
\sum_{l=1,2} \left[ \delta B_l, B_l^{\dagger}\right]+\delta II^{\dagger}-J^{\dagger}\delta J=0
\label{fermionicADHM}
\end{eqnarray}
around the fixed points. Denote the fixed point ADHM data corresponding to the Young diagrams $(Y_1, Y_2)$ as
${\cal A}_{Y_1,Y_2}$. The function $Z_k \left( a, \epsilon_1, \epsilon_2 \right)$ can be completely determined
in terms of above data:
\begin{eqnarray}
Z_k \left(a, \epsilon_1, \epsilon_2 \right)= \left. \sum_{Y_1,
Y_2} \frac{1}{\det {\cal L}_{{\tilde x}'}} \, \, \right|_{{\cal
A}_{Y_1,Y_2}}. \label{localization}
\end{eqnarray}
The reason of appearance of $1/\det $ instead of $1/\sqrt{\det}$
as in Eq. (\ref{normalbundleint}) is due to our convention to
consider complex tangent spaces and linear operators acting  on
them instead of their real forms.

Though the deformed superinstanton action $d_{{\tilde x}'}{\tilde
\omega}$ tends to the initial one as $\epsilon_1, \epsilon_2
\rightarrow 0$, it is not true for the $Z_k(a, \epsilon_1,
\epsilon_2)$ which is highly singular at this limit. In his recent
paper N.Nekrasov \cite{N} brings some field theoretical arguments
to come to the remarkable conclusion that the generating function
\begin{equation}
Z\left(q, a, \epsilon_1, \epsilon_2 \right)\equiv 1+\sum_{k=1}^{\infty} Z_k\left(a, \epsilon_1, \epsilon_2
\right)q^k \label{generatingZ}
\end{equation}
can be represented as
\begin{equation}
Z\left(q, a, \epsilon_1, \epsilon_2
\right)=\exp{-\frac{1}{\epsilon_1 \epsilon_2}{\cal F}\left(q, a,
\epsilon_1, \epsilon_2 \right)}, \label{singularitystructure}
\end{equation}
where ${\cal F}$ is regular at $\epsilon_1, \epsilon_2 =0$ and
that the SW coefficients ${\cal F}_k$ are nothing else than the
coefficients of the Taylor expansion of ${\cal F}\left(q, a,
\epsilon_1, \epsilon_2 \right)$:
\begin{equation}
{\cal F}\left(q, a, \epsilon_1, \epsilon_2 \right)=\sum_{k=1}^{\infty}{\cal F}_k\left(a, \epsilon_1, \epsilon_2
\right)q^k \label{TaylorF}
\end{equation}
at $\epsilon_1, \epsilon_2 =0$. More precisely\footnote{The factor
$2^{2k-2}$ is needed to have the normalization adopted in
\cite{dorey1}.}
\begin{equation}
{\cal F}_k a^{2-4k} =2^{2k-2}{\cal F}_k\left(a, \epsilon_1, \epsilon_2 \right).
\end{equation}
We hope to return to this point in a future publication and to
present an intrinsic explanation of the singularity structure
(\ref{singularitystructure}) using directly the definition of
$Z\left(q, a, \epsilon_1, \epsilon_2 \right)$.

\setcounter{equation}{0}
\section{Low charge instanton calculations}
\renewcommand{\theequation}{4.\arabic{equation}}

In this section we carry out explicit calculations for instanton charges up to four. For simplicity we rescale
ADHM data and set $\zeta=1$.
\begin{itemize}
\item {\bf One-instantons} \end{itemize}
This case is almost trivial. We have two pairs of Young diagrams\footnote{We always note only nonzero $\nu$'s.}: \\
{\bf a)} $Y_1=\left\{\nu_{1,1}=1\right\}$, $Y_2=\left\{\emptyset\right\}$; \\
{\bf b)} $Y_1=\left\{\emptyset\right\}$,
$Y_2=\left\{\nu_{1,1}=1\right\}$. In the case a) $B_1=B_2=0$,
$I=\left(\begin{array}{cc}1 & 0 \end{array}\right)$,
$\Phi_1=-a_1$. The tangent space vectors are given by
\begin{eqnarray}
\delta B_l=\left(\delta B_{l,11}\right); \, \, \,
\delta I=\left(
\begin{array}{cc} 0 & \delta I_{12}
\end{array}\right);
\, \, \,
\delta J=\left(
\begin{array}{c}
\delta J_{11} \\
0
\end{array}\right).
\end{eqnarray}
The dimension of the tangent space is $2\times 4=8$ as it should. Taking into account (\ref{xtildeaction}) one
easily calculates the determinant
\begin{equation}
\det {\cal L}_{{\tilde x}'}=\epsilon_1 \epsilon_2 a_{21}
\left(a_{12}+\epsilon \right),
\end{equation}
where $a_{\lambda \mu}\equiv a_{\lambda}-a_{\mu}; \lambda , \mu
=1,2$. There is no need to carry out calculation for the case b)
because for interchanged $Y_1$ and $Y_2$ one obtains the same
determinant with interchanged $a_1$ and $a_2$. Note also that
simultaneous transposition of both Young diagrams gives rise to a
determinant with interchanged $\epsilon_1 \leftrightarrow
\epsilon_2$. Below we will explicitly mention only one pair of
such symmetry related diagrams, but of course we will take all of
them into account in final expressions of $Z_k$ Eq.
(\ref{localization}). Thus for the $1$-instantons
\begin{equation}
Z_1\left(a, \epsilon_1, \epsilon_2 \right)=\left(\epsilon_1 \epsilon_2 a_{21} \left(a_{12}+\epsilon
\right)\right)^{-1}+\left(\epsilon_1 \epsilon_2 a_{12} \left(a_{21}+\epsilon \right)\right)^{-1}.
\end{equation}
From Eq. (\ref{generatingZ})-(\ref{TaylorF})
\begin{equation}
{\cal F}_1\left(a, \epsilon_1, \epsilon_2 \right)=-\epsilon_1
\epsilon_2 Z_1 =\frac{2}{a_{12}^2-\epsilon^2}.
\end{equation}
Taking the limit $\epsilon_{1,2}\rightarrow 0$ one obtains
\[
{\cal F}_1=1/2.
\]
\begin{itemize}
\item {\bf 2-instantons} \end{itemize} There are two basic cases.
\\
{\bf a)} $Y_1=\left\{\nu_{1,1}=2)\right\}$,
$Y_2=\left\{\emptyset\right\}$;
$\Phi_1=-a_1$, $\Phi_2=-a_1-\epsilon_1$. \\
The fixed point ADHM data:
\begin{eqnarray}
B_1=\left(\begin{array}{cc}
0 & 0 \\ 1 & 0
\end{array}\right); \, \, \, \,
B_2=0; \, \, \, \, I=\left(\begin{array}{cc} \sqrt{2} & 0 \\ 0 & 0
\end{array}\right).
\end{eqnarray}
The tangent space:
\begin{eqnarray}
\delta B_1 &=& \left(\begin{array}{cc} \delta B_{1,11} & \delta B_{1,12} \\ 0 & \delta B_{1,11}
\end{array}\right); \, \, \, \,
\delta B_2 =\left(\begin{array}{cc} \delta B_{2,11} & 0 \\ \delta B_{2,21} & \delta B_{2,11}
\end{array}\right); \nonumber \\
\delta I &=&\left(\begin{array}{cc} 0 & \delta I_{12} \\ 0 & \delta I_{22}
\end{array}\right); \, \, \, \, \, \, \, \, \, \, \, \, \, \, \,
\delta J=\left(\begin{array}{cc} 0 & 0 \\ \delta J_{21} & \delta J_{22}
\end{array}\right).
\end{eqnarray}
The determinant:
\begin{equation}
\det {\cal L}_{{\tilde x}'}=2 \epsilon_1^2 \epsilon_2
\epsilon_{21} a_{21} \left(a_{21}-\epsilon_1 \right)
\left(a_{12}+\epsilon \right) \left(a_{12}+2 \epsilon_1
+\epsilon_2 \right).
\end{equation}
{\bf b)} $Y_1=\left\{\nu_{1,1}=1)\right\}$, $Y_2=\left\{\nu_{2,1}=1\right\}$; $\Phi_1=-a_1$, $\Phi_2=-a_2$. \\
Fixed point:
\begin{eqnarray}
B_1=B_2=0; \, \, \, \, I=\left(\begin{array}{cc} 1 & 0 \\ 0 & 1
\end{array}\right).
\end{eqnarray}
Tangent space:
\begin{eqnarray}
\delta B_1 &=& \left(\begin{array}{cc} \delta B_{1,11} & \delta B_{1,12} \\ \delta B_{1,21} & \delta B_{1,22}
\end{array}\right); \, \, \, \,
\delta B_2 =\left(\begin{array}{cc} \delta B_{2,11} & \delta B_{2,12} \\ \delta B_{2,21} & \delta B_{2,22}
\end{array}\right); \nonumber \\ \delta I &=& 0; \, \, \, \delta J=0.
\end{eqnarray}
Determinant:
\begin{equation}
\det {\cal L}_{{\tilde x}'}=\epsilon_1^2 \epsilon_2^2
\left(a_{12}^2-\epsilon_1^2 \right)\left(a_{12}^2-\epsilon_2^2
\right).
\end{equation}
Using above data one first calculates $Z_2$ and then
\[ {\cal F}_2\left(a, \epsilon_1,\epsilon_2\right)=-\epsilon_1 \epsilon_2 \left( Z_2-\frac{1}{2} Z_1^2\right). \]
The final result reads:
\begin{equation}
{\cal F}_2\left(a, \epsilon_1, \epsilon_2\right)=\frac{20 a^2 + 7 \epsilon_1^2 + 16 \epsilon_1 \epsilon_2 + 7
\epsilon_2^2}{\left(4a^2 - \epsilon^2 \right)^2  \left(4a^2 - \left( 2\epsilon_1 + \epsilon_2\right)^2 \right)
 \left(4a^2 - \left(\epsilon_1 + 2\epsilon_2 \right)^2 \right)}.
\end{equation}
Now let us take the limit $\epsilon_{1,2}\rightarrow 0$:
\[
{\cal F}_2=\frac{5}{2^{4}}.
\]

\begin{itemize}
\item {\bf 3-instantons} \end{itemize} There are three basic
cases. \\
{\bf a)} $Y_1=\left\{\nu_{1,1}=3\right\}$,
$Y_2=\left\{\emptyset\right\}$; $\Phi_1=-a_1$,
$\Phi_2=-a_1-\epsilon_1$,
$\Phi_3=-a_1-2\epsilon_1$. \\
Fixed point:
\begin{eqnarray}
B_1=\left(\begin{array}{ccc} 0 & 0 & 0\\ \sqrt{2} & 0 & 0 \\ 0 & 1 & 0
\end{array}\right); \, \, \, \,
B_2=0; \, \, \, \, I=\left(\begin{array}{cc} \sqrt{3} & 0 \\ 0 & 0 \\0 & 0
\end{array}\right).
\end{eqnarray}
Tangent space:
\begin{eqnarray}
\delta B_1 &=& \left(\begin{array}{ccc} \delta B_{1,11} & \delta B_{1,12} & \delta B_{1,13}\\ 0 & \delta
B_{1,11} & \frac{1}{\sqrt{2}}\delta B_{1,12} \\0 & 0 & \delta B_{1,11}
\end{array}\right); \, \, \, \,
\delta B_2 =\left(\begin{array}{ccc} \delta B_{2,11} & 0 & 0 \\ \delta B_{2,21} & \delta B_{2,11} & 0 \\ \delta
B_{2,31} & \frac{1}{\sqrt{2}}\delta B_{2,21} & \delta B_{2,11}
\end{array}\right); \nonumber \\
\delta I &=&\left(\begin{array}{cc} 0 & \delta I_{12} \\ 0 & \delta I_{22} \\ 0 & \delta I_{32}
\end{array}\right); \, \, \, \, \, \, \, \, \, \, \, \, \, \, \,
\delta J=\left(\begin{array}{ccc} 0 & 0 & 0 \\ \delta J_{21} & \delta J_{22} & \delta J_{23}
\end{array}\right).
\end{eqnarray}
The determinant:
\begin{eqnarray}
\det {\cal L}_{{\tilde x}'}=6 \epsilon_1^3 \epsilon_2
\epsilon_{21} \left(\epsilon_2-2\epsilon_1\right) a_{21}
\left(a_{21}-\epsilon_1 \right) \left(a_{21}-2\epsilon_1 \right) \nonumber \\
\times \left(a_{12}+\epsilon \right) \left(a_{12}+2 \epsilon_1 +\epsilon_2 \right) \left(a_{12}+3 \epsilon_1
+\epsilon_2 \right).
\end{eqnarray}
{\bf b)} $Y_1=\left\{\nu_{1,1}=2, \nu_{1,2}=1 \right\}$,
$Y_2=\left\{\emptyset\right\}$; $\Phi_1=-a_1$,
$\Phi_2=-a_1-\epsilon_1$, $\Phi_3=-a_1-\epsilon_1-\epsilon_2$. \\
Fixed point:
\begin{eqnarray}
B_1=\left(\begin{array}{ccc} 0 & 0 & 0\\ 1 & 0 & 0 \\ 0 & 0 & 0
\end{array}\right); \, \, \, \,
B_2=\left(\begin{array}{ccc} 0 & 0 & 0\\ 0 & 0 & 0 \\ 1 & 0 & 0
\end{array}\right);  \, \, \, \, I=\left(\begin{array}{cc} \sqrt{3} & 0 \\ 0 & 0 \\0 & 0
\end{array}\right).
\end{eqnarray}
Tangent space:
\begin{eqnarray}
\delta B_1 &=& \left(\begin{array}{ccc} \delta B_{1,11} & 0 & 0 \\ 0 & \delta B_{1,22} & \delta B_{2,22}-\delta
B_{2,11}\\0 & \delta B_{1,32} & 2 \delta B_{1,11}-\delta B_{1,22}
\end{array}\right); \nonumber \\
\delta B_2 &=&\left(\begin{array}{ccc} \delta B_{2,11} & 0 & 0 \\ 0 & \delta B_{2,22} & \delta B_{2,23} \\
0 & \delta B_{1,11}-\delta B_{1,22} & 2 \delta B_{211}-\delta B_{2,22}
\end{array}\right); \nonumber \\
\delta I &=& \left(\begin{array}{cc} 0 & \delta I_{12} \\ 0 & \delta I_{22} \\ 0 & \delta I_{32}
\end{array}\right); \, \, \, \, \, \, \, \, \,
\delta J=\left(\begin{array}{ccc} 0 & 0 & 0 \\ \delta J_{21} & \delta J_{22} & \delta J_{23}
\end{array}\right).
\end{eqnarray}
Determinant:
\begin{eqnarray}
\det {\cal L}_{{\tilde x}'}=\epsilon_1^2 \epsilon_2^2 \left(2
\epsilon_1-\epsilon_2\right)\left(2
\epsilon_2-\epsilon_1\right)a_{21}\left(a_{21}-\epsilon_1 \right)\left(a_{21}-\epsilon_2 \right) \nonumber \\
\times \left(a_{12}+\epsilon \right)\left(a_{12}+2 \epsilon_1 +\epsilon_2 \right)\left(a_{12}+2 \epsilon_2
+\epsilon_1 \right).
\end{eqnarray}
{\bf c)} $Y_1=\left\{\nu_{1,1}=1, \nu_{1,2}=1 \right\}$,
$Y_2=\left\{\nu_{2,1}=1\right\}$; $\Phi_1=-a_1$,
$\Phi_2=-a_1-\epsilon_2$, $\Phi_3=-a_2$. \\
Fixed point:
\begin{eqnarray}
B_1=0; \, \, \, \, B_2=\left(\begin{array}{ccc} 0 & 0 & 0\\ 1 & 0 & 0 \\ 0 & 0 & 0
\end{array}\right);  \, \, \, \, I=\left(\begin{array}{cc} \sqrt{2} & 0 \\ 0 & 0 \\0 & 1
\end{array}\right).
\end{eqnarray}
Tangent space:
\begin{eqnarray}
\delta B_1 &=& \left(\begin{array}{ccc} \delta B_{1,11} & 0 & 0 \\
\delta B_{1,21}  & \delta B_{1,11} & \delta B_{1,23} \\ \delta
B_{1,31} & \delta B_{1,32} & \delta B_{1,33}
\end{array}\right); \nonumber \\
\delta B_2 &=&\left(\begin{array}{ccc} \delta B_{2,11} & \delta B_{2,12} & \delta B_{2,13} \\
0 & \delta B_{2,11} & \delta B_{2,23} \\
 0 & \delta B_{2,32} & \delta B_{2,33}
\end{array}\right); \nonumber \\
\delta I &=& \left(\begin{array}{cc} 0 & \delta B_{2,23} \\ 0 & 0
\\ 0 & 0
\end{array}\right); \, \, \, \, \, \, \, \, \,
\delta J=\left(\begin{array}{ccc} 0 & 0 & 0 \\ -\delta B_{1,32} &
0 & 0
\end{array}\right).
\end{eqnarray}
Determinant:
\begin{eqnarray}
\det {\cal L}_{{\tilde x}'}=2 \epsilon_1^2 \epsilon_2^3
\epsilon_{12} a_{21}
\left(a_{21}+\epsilon_1-\epsilon_2 \right)\left(a_{12}+\epsilon_1 \right) \left(a_{21}+\epsilon_2 \right)\nonumber \\
\times \left(a_{12}+\epsilon \right)\left(a_{12}+2 \epsilon_2 \right).
\end{eqnarray}
Using these data we have calculated $Z_3$ and then
\[ {\cal F}_3\left(a, \epsilon_1,\epsilon_2\right)=-\epsilon_1 \epsilon_2 \left(Z_3 +
\frac{1}{3} Z_1^3 - Z_1 Z_2 \right). \]
Here is the final result:
\begin{eqnarray}
{\cal F}_3\left(a, \epsilon_1,\epsilon_2\right)= \hspace{6 cm} \\ \frac{16 \left(144 a^4 + 29 \epsilon_1^4 + 154
\epsilon_1^3 \epsilon_2 + 258 \epsilon_1^2 \epsilon_2^2 +154 \epsilon_1 \epsilon_2^3 + 29 \epsilon_2^4 + 8 a^2
\left(29 \epsilon_1^2 + 71 \epsilon_1 \epsilon_2 + 29 \epsilon_2^2 \right) \right)}{3\left(4a^2 - \epsilon^2
\right)^3 \left(4a^2 -\left( 2 \epsilon_1  + \epsilon_2\right)^2 \right) \left(4a^2 -\left( 3\epsilon_1
+\epsilon_2\right)^2 \right)\left( 4a^2 -\left( \epsilon_1 + 2\epsilon_2\right)^2 \right) \left(4a^2 -\left(
\epsilon_1 + 3\epsilon_2\right)^2 \right)}. \nonumber
\end{eqnarray}
Now let us take the limit $\epsilon_{1,2}\rightarrow 0$:
\[
{\cal F}_3=\frac{3}{4}.
\]
\begin{itemize}
\item {\bf 4-instantons}
\end{itemize}
Now we need to investigate 7 different cases.\\
{\bf a)} $Y_1=\left\{\nu_{1,1}=4\right\}$,
$Y_2=\left\{\emptyset\right\}$; $\Phi_1=-a_1$,
$\Phi_2=-a_1-\epsilon_1$,
$\Phi_3=-a_1-2\epsilon_1$, $\Phi_4=-a_1-3\epsilon_1$. \\
Fixed point:
\begin{eqnarray}
B_1=\left(\begin{array}{cccc} 0 & 0 & 0 & 0 \\ \sqrt{3} & 0 & 0 & 0 \\ 0 & \sqrt{2} & 0 & 0 \\
0 & 0 & 1 & 0
\end{array}\right); \, \, \, \,
B_2=0; \, \, \, \, I=\left(\begin{array}{cc} 2 & 0 \\ 0 & 0 \\0 & 0 \\ 0 & 0
\end{array}\right).
\end{eqnarray}
Tangent space:
\begin{eqnarray}
\delta B_1 &=& \left(\begin{array}{cccc} \delta B_{1,11} & \delta B_{1,12} & \delta B_{1,13} & \delta B_{1,14}\\
0 & \delta B_{1,11} & \sqrt{\frac{2}{3}}\delta B_{1,12} &
\frac{1}{\sqrt{3}} \delta B_{1,13} \\ 0 & 0 & \delta B_{1,11} &
\frac{1}{\sqrt{3}} \delta B_{1,12} \\ 0 & 0 & 0 & \delta B_{1,11}
\end{array}\right); \nonumber \\
\delta B_2 &=& \left(\begin{array}{cccc} \delta B_{2,11} & 0 & 0 & 0 \\ \delta B_{2,21} & \delta B_{211} & 0 & 0 \\
\delta B_{2,31} & \sqrt{\frac{2}{3}} \delta B_{2,21} & \delta B_{2,11} & 0 \\
\delta B_{2,41} & \frac{1}{\sqrt{3}} \delta B_{2,31} & \frac{1}{\sqrt{3}} \delta B_{2,21} & \delta B_{2,11}
\end{array}\right); \nonumber \\
\delta I &=& \left(\begin{array}{cc} 0 & \delta I_{12} \\ 0 & \delta I_{22} \\ 0 & \delta I_{32} \\ 0 & \delta
I_{42}
\end{array}\right); \, \, \, \, \, \, \,
\delta J=\left(\begin{array}{cccc} 0 & 0 & 0 & 0 \\ \delta J_{21} & \delta J_{22} & \delta J_{23} & \delta
J_{24}
\end{array}\right).
\end{eqnarray}
The determinant:
\begin{eqnarray}
\det {\cal L}_{{\tilde x}'}=24 \epsilon_1^4 \prod_{j=0}^3
\left(\epsilon_2-j\epsilon_1 \right) \left(a_{21}-j\epsilon_1
\right) \left(a_{12}+ \epsilon +j \epsilon_1 \right).
\end{eqnarray}
{\bf b)} $Y_1=\left\{\nu_{1,1}=3, \nu_{1,2}=1 \right\}$,
$Y_2=\left\{\emptyset\right\}$; $\Phi_1=-a_1$,
$\Phi_2=-a_1-\epsilon_1$,
$\Phi_3=-a_1-2\epsilon_1$, $\Phi_4=-a_1-\epsilon_2$. \\
Fixed point:
\begin{eqnarray}
B_1=\left(\begin{array}{cccc} 0 & 0 & 0 & 0 \\ \sqrt{2} & 0 & 0 & 0 \\ 0 & 1 & 0 & 0 \\
0 & 0 & 0 & 0
\end{array}\right); \, \, \, \,
B_2=\left(\begin{array}{cccc} 0 & 0 & 0 & 0 \\ 0 & 0 & 0 & 0 \\ 0 & 0 & 0 & 0 \\
1 & 0 & 0 & 0
\end{array}\right);
\, \, \, \, I=\left(\begin{array}{cc} 2 & 0 \\ 0 & 0 \\0 & 0 \\ 0 & 0
\end{array}\right).
\end{eqnarray}
Tangent space:
\begin{eqnarray}
\delta B_1 &=& \left(\begin{array}{cccc} \delta B_{1,11} & \delta B_{1,12} & 0 & 0 \\
0 & \delta B_{1,22} & \sqrt{2}\delta B_{1,12} & \sqrt{2} \left(\delta B_{2,22} -\delta B_{2,11} \right) \\ 0 & 0
& \delta B_{1,22} & \delta B_{1,34} \\ 0 & 0 & \delta_{1,43} & 3 \delta B_{1,11}-2 \delta B_{1,22}
\end{array}\right); \nonumber \\
\delta B_2 &=& \left(\begin{array}{cccc} \delta B_{2,11} & 0 & 0 & 0 \\ \delta B_{1,34} & \delta B_{2,22} & 0 & 0 \\
0 & \sqrt{2} \delta B_{1,34} & \delta B_{2,22} & \delta B_{2,34} \\
0 & \sqrt{2}\left( \delta B_{1,11}-\delta B_{1,22}\right) & -\delta B_{1,12} & 3 \delta B_{2,11}-2 \delta
B_{2,22}
\end{array}\right); \nonumber \\
\delta I &=& \left(\begin{array}{cc} 0 & \delta I_{12} \\ 0 & \delta I_{22} \\ 0 & \delta I_{32} \\ 0 & \delta
I_{42}
\end{array}\right); \, \, \, \, \, \, \,
\delta J=\left(\begin{array}{cccc} 0 & 0 & 0 & 0 \\ \delta J_{21} & \delta J_{22} & \delta J_{23} & \delta
J_{24}
\end{array}\right).
\end{eqnarray}
The determinant:
\begin{eqnarray}
\det {\cal L}_{{\tilde x}'}=4 \epsilon_1^3 \epsilon_2^2
\epsilon_{21}^2 \left(3 \epsilon_1 -\epsilon_2 \right)a_{21}
\left(a_{21}-\epsilon_1 \right) \left(a_{21}-2 \epsilon_1
\right)\left(a_{21}-\epsilon_2
\right)\left(a_{12}+\epsilon \right) \nonumber \\
\times \left(a_{12}+2\epsilon_1 +\epsilon_2 \right)\left(a_{12}+3\epsilon_1 +\epsilon_2
\right)\left(a_{12}+\epsilon_1 +2\epsilon_2 \right).
\end{eqnarray}
{\bf c)} $Y_1=\left\{\nu_{1,1}=2, \nu_{1,2}=2 \right\}$,
$Y_2=\left\{\emptyset\right\}$; $\Phi_1=-a_1$,
$\Phi_2=-a_1-\epsilon_1$,
$\Phi_3=-a_1-\epsilon_2$, $\Phi_4=-a_1-\epsilon$. \\
Fixed point:
\begin{eqnarray}
B_1=\left(\begin{array}{cccc} 0 & 0 & 0 & 0 \\ \sqrt{\frac{3}{2}} & 0 & 0 & 0 \\ 0 & 0 & 0 & 0 \\
0 & 0 & \frac{1}{\sqrt{2}} & 0
\end{array}\right); \, \, \, \,
B_2=\left(\begin{array}{cccc} 0 & 0 & 0 & 0 \\ 0 & 0 & 0 & 0  \\ \sqrt{\frac{3}{2}} & 0 & 0 & 0 \\
0 & \frac{1}{\sqrt{2}} & 0 & 0
\end{array}\right);
\, \, \, \, I=\left(\begin{array}{cc} 2 & 0 \\ 0 & 0 \\0 & 0 \\ 0 & 0
\end{array}\right).
\end{eqnarray}
Tangent space:
\begin{eqnarray}
\delta B_1 &=& \left(\begin{array}{cccc} \delta B_{1,11} & \delta B_{1,12} & 0 & 0 \\
0 & \delta B_{1,11} & 0 & 0 \\ \delta B_{1,31} & \delta B_{1,32} & \delta B_{1,11} & \sqrt{3}\delta B_{1,12}
\\ 0 & \sqrt{3} \delta B_{1,31} & 0 & \delta B_{1,11}
\end{array}\right); \nonumber \\
\delta B_2 &=& \left(\begin{array}{cccc} \delta B_{2,11} & 0 & \delta B_{2,13} & 0 \\ \delta B_{2,21} & \delta B_{2,11} & \delta B_{2,23} & \sqrt{3} \delta B_{2,13} \\
0 & 0 & \delta B_{2,11} & 0 \\
0 & 0 & \sqrt{3} \delta B_{2,21} & \delta B_{2,11}
\end{array}\right); \nonumber \\
\delta I &=& \left(\begin{array}{cc} 0 & \delta I_{12} \\ 0 & \delta I_{22} \\ 0 & \delta I_{32} \\ 0 & \delta
I_{42}
\end{array}\right); \, \, \, \, \, \, \,
\delta J=\left(\begin{array}{cccc} 0 & 0 & 0 & 0 \\ \delta J_{21} & \delta J_{22} & \delta J_{23} & \delta
J_{24}
\end{array}\right).
\end{eqnarray}
The determinant:
\begin{eqnarray}
\det {\cal L}_{{\tilde x}'}=4 \epsilon_1^2 \epsilon_2^2
\epsilon_{12}^2 \left(2 \epsilon_1 -\epsilon_2 \right)\left(2
\epsilon_2 -\epsilon_1 \right)a_{21} \left(a_{21}-\epsilon_1
\right) \left(a_{21}-\epsilon_2
\right)\left(a_{12}+\epsilon \right)^2 \nonumber \\
\times \left(a_{12}+2\epsilon_1 +\epsilon_2 \right)\left(a_{12}+2\epsilon_2 +\epsilon_1
\right)\left(a_{12}+2\epsilon \right).
\end{eqnarray}
{\bf d)} $Y_1=\left\{\nu_{1,1}=1 \right\}$,
$Y_2=\left\{\nu_{2,1}=3 \right\}$; $\Phi_1=-a_1$, $\Phi_2=-a_2$,
$\Phi_3=-a_2-\epsilon_1$, $\Phi_4=-a_2-2\epsilon_1$. \\
Fixed point:
\begin{eqnarray}
B_1=\left(\begin{array}{cccc} 0 & 0 & 0 & 0 \\ 0 & 0 & 0 & 0 \\ 0 & \sqrt{2} & 0 & 0 \\
0 & 0 & 1 & 0
\end{array}\right); \, \, \, \, B_2=0;
 \, \, \, \, I=\left(\begin{array}{cc} 1 & 0 \\ 0 & \sqrt{3} \\0 & 0 \\ 0 & 0
\end{array}\right).
\end{eqnarray}
Tangent space:
\begin{eqnarray}
\delta B_1 &=& \left(\begin{array}{cccc} \delta B_{1,11} & 0 & 0 & \delta B_{1,14} \\
\delta B_{1,21} & \delta B_{1,22} & \delta B_{1,23} & \delta B_{1,24} \\ \delta B_{1,31} & 0 & \delta B_{1,22} &
\frac{1}{\sqrt{2}} \delta B_{1,23}
\\ \delta B_{1,41} & 0 & 0 & \delta B_{1,22}
\end{array}\right); \nonumber \\
\delta B_2 &=& \left(\begin{array}{cccc} \delta B_{2,11} & \delta B_{2,12} & \delta B_{2,13} & \delta B_{2,14} \\ 0 & \delta B_{2,22} & 0 & 0 \\
0 & \delta B_{2,32} & \delta B_{2,22} & 0 \\
\delta B_{2,41} & \delta B_{2,42} & \frac{1}{\sqrt{2}} \delta B_{2,32} & \delta B_{2,22}
\end{array}\right); \nonumber \\
\delta I &=& \left(\begin{array}{cc} 0 & 0 \\ \sqrt{2}\delta B_{1,31} & 0 \\ \delta B_{1,41} & 0 \\
0 & 0
\end{array}\right); \, \, \, \, \, \, \,
\delta J=\left(\begin{array}{cccc} 0 & \sqrt{2} \delta B_{2,13} & \delta B_{2,14} & 0 \\ 0 & 0 & 0 & 0
\end{array}\right).
\end{eqnarray}
The determinant:
\begin{eqnarray}
\det {\cal L}_{{\tilde x}'}=6\epsilon_1^4 \epsilon_2^2
\epsilon_{21} \left(\epsilon_2 -2\epsilon_1 \right) a_{12}
\left(a_{21}+3\epsilon_1 \right) \left(a_{12}^2-\epsilon_1^2
\right)\left(a_{21}+\epsilon_2 \right) \nonumber \\
\times \left(a_{21}+\epsilon \right)\left(a_{21}+2\epsilon_1 +\epsilon_2 \right)\left(a_{12}-2\epsilon_1
+\epsilon_2 \right).
\end{eqnarray}
{\bf e)} $Y_1=\left\{\nu_{1,1}=1 \right\}$,
$Y_2=\left\{\nu_{2,1}=2, \nu_{2,2}=1 \right\}$; $\Phi_1=-a_1$,
$\Phi_2=-a_2$,
$\Phi_3=-a_2-\epsilon_1$, $\Phi_4=-a_2-\epsilon_2$. \\
Fixed point:
\begin{eqnarray}
B_1=\left(\begin{array}{cccc} 0 & 0 & 0 & 0 \\ 0 & 0 & 0 & 0 \\ 0 & 1 & 0 & 0 \\
0 & 0 & 0 & 0
\end{array}\right); \, \, \, \,
B_2=\left(\begin{array}{cccc} 0 & 0 & 0 & 0 \\ 0 & 0 & 0 & 0 \\ 0 & 0 & 0 & 0 \\
0 & 1 & 0 & 0
\end{array}\right);
 \, \, \, \, I=\left(\begin{array}{cc} 1 & 0 \\ 0 & \sqrt{3} \\0 & 0 \\ 0 & 0
\end{array}\right).
\end{eqnarray}
Tangent space:
\begin{eqnarray}
\delta B_1 &=& \left(\begin{array}{cccc} \delta B_{1,11} & 0 & \delta B_{1,13} & \delta B_{1,14} \\
0 & \delta B_{1,22} & 0 & 0 \\ \delta B_{1,31} & 0 & \delta B_{1,33} & \delta B_{2,33}-\delta B_{2,22}
\\ \delta B_{1,41} & 0 & \delta B_{1,43} & 2\delta B_{1,22}-\delta B_{1,33}
\end{array}\right); \nonumber \\
\delta B_2 &=& \left(\begin{array}{cccc} \delta B_{2,11} & 0 & \delta B_{2,13} & \delta B_{2,14} \\ 0 & \delta B_{2,22} & 0 & 0 \\
\delta B_{2,31} & 0 & \delta B_{2,33} & \delta B_{2,34} \\
\delta B_{2,41} & 0 & \delta B_{1,22}-\delta B_{1,33} & 2\delta B_{2,22}-\delta B_{2,33}
\end{array}\right); \nonumber \\
\delta I &=& \left(\begin{array}{cc} 0 & 0 \\ \delta B_{1,31}+\delta B_{2,41} & 0 \\ 0 & 0 \\
0 & 0
\end{array}\right); \, \, \, \, \, \, \,
\delta J=\left(\begin{array}{cccc} 0 & \delta B_{2,13}-\delta B_{1,14} & 0 & 0 \\ 0 & 0 & 0 & 0
\end{array}\right).
\end{eqnarray}
The determinant:
\begin{eqnarray}
\det {\cal L}_{{\tilde x}'}=\epsilon_1^3 \epsilon_2^3
\left(2\epsilon_1 -\epsilon_2 \right) \left(2\epsilon_2
-\epsilon_1 \right)a_{12}^2 \left(a_{21}+2\epsilon_1 \right)
\left(a_{21}+2\epsilon_2 \right)\left(a_{21}+\epsilon
\right)^2\left(a_{12}^2-\epsilon_{12}^2 \right).
\end{eqnarray}
{\bf f)} $Y_1=\left\{\nu_{1,1}=2 \right\}$,
$Y_2=\left\{\nu_{2,1}=2 \right\}$; $\Phi_1=-a_1$,
$\Phi_2=-a_1-\epsilon_1$,
$\Phi_3=-a_2$, $\Phi_4=-a_2-\epsilon_1$. \\
Fixed point:
\begin{eqnarray}
B_1=\left(\begin{array}{cccc} 0 & 0 & 0 & 0 \\ 1 & 0 & 0 & 0 \\ 0 & 0 & 0 & 0 \\
0 & 0 & 1 & 0
\end{array}\right); \, \, \, \,
B_2=0; \, \, \, \, I=\left(\begin{array}{cc} \sqrt{2} & 0 \\ 0 & 0 \\ 0 & \sqrt{2} \\0 & 0
\end{array}\right).
\end{eqnarray}
Tangent space:
\begin{eqnarray}
\delta B_1 &=& \left(\begin{array}{cccc} \delta B_{1,11} & \delta B_{1,12} & \delta B_{1,13} & \delta B_{1,14} \\
0 & \delta B_{1,11} & 0 & \delta B_{1,13} \\
\delta B_{1,31} & \delta B_{1,32} & \delta B_{1,33} & \delta B_{1,34}
\\ 0 & \delta B_{1,31} & 0 & \delta B_{1,33}
\end{array}\right); \nonumber \\
\delta B_2 &=& \left(\begin{array}{cccc} \delta B_{2,11} & 0 & \delta B_{2,13} & 0 \\ \delta B_{2,21} & \delta B_{2,11} & \delta B_{2,23} & \delta B_{2,13} \\
\delta B_{2,31} & 0 & \delta B_{2,33} & 0 \\
\delta B_{2,41} & \delta B_{2,31} & \delta B_{2,43} & \delta B_{2,33}
\end{array}\right); \nonumber \\
\delta I &=& 0; \, \, \, \, \, \, \, \delta J= 0.
\end{eqnarray}
The determinant:
\begin{eqnarray}
\det {\cal L}_{{\tilde x}'}=4\epsilon_1^4 \epsilon_2^2
\epsilon_{12}^2 \left(a_{12}^2-\epsilon_1^2
\right)\left(a_{12}^2-4\epsilon_1^2
\right)\left(a_{12}^2-\epsilon_2^2
\right)\left(a_{12}^2-\epsilon_{12}^2 \right).
\end{eqnarray}
{\bf g)} $Y_1=\left\{\nu_{1,1}=2 \right\}$,
$Y_2=\left\{\nu_{2,1}=1, \nu_{2,2}=1 \right\}$; $\Phi_1=-a_1$,
$\Phi_2=-a_1-\epsilon_1$,
$\Phi_3=-a_2$, $\Phi_4=-a_2-\epsilon_2$. \\
Fixed point:
\begin{eqnarray}
B_1=\left(\begin{array}{cccc} 0 & 0 & 0 & 0 \\ 1 & 0 & 0 & 0 \\ 0 & 0 & 0 & 0 \\
0 & 0 & 0 & 0
\end{array}\right); \, \, \, \,
B_2=\left(\begin{array}{cccc} 0 & 0 & 0 & 0 \\ 0 & 0 & 0 & 0 \\ 0 & 0 & 0 & 0 \\
0 & 0 & 1 & 0
\end{array}\right); \, \, \, \, I=\left(\begin{array}{cc} \sqrt{2} & 0 \\ 0 & 0 \\ 0 & \sqrt{2} \\0 & 0
\end{array}\right).
\end{eqnarray}
Tangent space:
\begin{eqnarray}
\delta B_1 &=& \left(\begin{array}{cccc} \delta B_{1,11} & \delta B_{1,12} & \delta B_{1,13} & \delta B_{1,14} \\
0 & \delta B_{1,11} & \delta B_{1,23} & 0 \\
0 & 0 & \delta B_{1,33} & 0
\\ 0 & \delta B_{1,42} & \delta B_{1,43} & \delta B_{1,33}
\end{array}\right); \nonumber \\
\delta B_2 &=& \left(\begin{array}{cccc} \delta B_{2,11} & 0 & 0 & 0 \\ \delta B_{2,21} & \delta B_{2,11} & 0 & \delta B_{2,24} \\
\delta B_{2,31} & \delta B_{2,32} & \delta B_{2,33} & \delta B_{2,34} \\
\delta B_{2,41} & 0 & 0 & \delta B_{2,33}
\end{array}\right); \nonumber \\
\delta I &=& \left(
\begin{array}{cc}
0 & \frac{1}{\sqrt{2}}\delta B_{1,23} \\ 0 & 0 \\ \frac{1}{\sqrt{2}}\delta B_{2,41} & 0 \\ 0 & 0
\end{array}
\right); \, \, \, \, \, \, \,
\delta J= \left(
\begin{array}{cccc}
0 & 0 & -\frac{1}{\sqrt{2}}\delta B_{1,14} & 0 \\ \frac{1}{\sqrt{2}}\delta B_{2,32} & 0 & 0 & 0
\end{array}
\right);
\end{eqnarray}
The determinant:
\begin{eqnarray}
\det {\cal L}_{{\tilde x}'}=4\epsilon_1^3 \epsilon_2^3
\epsilon_{12}^2 a_{12}^2 \left(a_{12}^2-\epsilon^2
\right)\left(a_{12}-\epsilon_1 \right)\left(a_{12}+\epsilon_2
\right)\left(a_{12}+2\epsilon_1-\epsilon_2
\right)\left(a_{21}+2\epsilon_2-\epsilon_1 \right).
\end{eqnarray}
The expression for
\[
{\cal F}_4\left(a, \epsilon_1,\epsilon_2\right)= \epsilon_1 \epsilon_2 \left(Z_4 - Z_1 Z_3 - \frac{1}{2} Z_2^2 +
Z_1^2 Z_2 - \frac{1}{4} Z_1^4\right)
\]
is very lengthy to present here. We only note here that it is
indeed regular at $\epsilon_{1,2}\rightarrow 0$. Here are the
first nontrivial terms of its expansion:
\begin{equation}
{\cal
F}_4\left(a,\epsilon_1,\epsilon_2\right)=\frac{1469}{2^{15}a^{14}}+
\frac{18445}{2^{15}a^{16}}\left(\epsilon_1^2+\epsilon_2^2\right)
+\frac{15151}{2^{14}a^{16}}\epsilon_1 \epsilon_2 +\cdots ,
\end{equation}
Thus
\[
{\cal F}_4=\frac{1469}{2^{9}},
\]
which is as it should be.

Finally we quote a general formula for the determinant of the
vector field action on the tangent space of a generic critical
point at arbitrary instanton number and gauge group $SU(N)$. To
obtain this formula we closely follow the line of arguments
presented in \cite{Nak}, Section 5.2, where the characters of the
torus action around fixed points are calculated. For our purposes
we need to calculate the character of the representation of the
group
\begin{equation}
U(1)^{N-1}\times U(1)^2 \label{group}
\end{equation}
(the first factor is the Cartan subgroup of the group $SU(N)$ and
the second is the 2-torus acting in space-time and on ADHM data)
in the tangent space at the fixed point specified by the Young
diagrams $Y_1, \dots, Y_N$. The result reads (cf. with the formula
in proposition 5.8 page 67 of Nakajima's book \cite{Nak}):
\begin{eqnarray}
\chi \left(a_l, \epsilon_1 , \epsilon_2\right)=\sum_{\lambda ,\mu
=1}^N T_{a_{\mu}} T_{a_{\lambda}}^{-1} \left(\sum_{s\in
Y_{\lambda}} T_1^{-h_{\lambda} (s)} T_2^{1+v_{\mu} (s)}+
\sum_{s'\in Y_{\mu}} T_1^{1+h_{\mu} (s')} T_2^{-v_{\lambda}
(s')}\right), \label{character}
\end{eqnarray}
where $h_{\lambda} (s)=\nu_{\lambda ,i} -j$, $v_{\lambda}
(s)=\nu'_{\lambda ,j} -i$ if the box $s$ is located on the $i$-th
row and the $j$-th column of a Young diagram. It is assumed that
$\nu_{\lambda ,i}$, $\nu'_{\lambda ,j}$ are defined for arbitrary
positive integers $i$, $j$. For $i > \nu'_{\lambda ,1}$ and $j>
\nu_{\lambda ,1}$ by definition they are identically zero. In
(\ref{character}) $T_{a_{\lambda}} \equiv \exp ia_{\lambda} $,
$T_{l} \equiv \exp i\epsilon_l$ are elements of respective $U(1)$
factors of the group (\ref{group}) taken in the fundamental
representations. A term of the form $T_{a_{\mu}}
T_{a_{\lambda}}^{-1} T_1^m T_2^n$ in Eq. (\ref{character})
indicates that the tangent space includes a (complex) one
dimensional invariant subspace of our deformed vector field
${\tilde x}'$ action with eigenvalue $a_{\mu}-a_{\nu}+m \epsilon_1
+n \epsilon_2$. Multiplying all these eigenvalues for the
determinant of ${\tilde x}'$ action we find\footnote{A more
complicated formula for the (inverse ) determinant is presented
also in \cite{N} Eq. (3.20) which has a discrepancy as compared to
our formula (\ref{generaldet}). We have checked that in all cases
explicitly described in this section the formula
(\ref{generaldet}) gives correct results.}
\begin{eqnarray}
\left. \det {\cal L}_{{\tilde x}'} \right|_{\left(Y_1, \cdots,
Y_N\right)}= \prod_{\lambda ,\mu =1}^N  \left( \prod_{s\in
Y_{\lambda}}\left( a_{\mu \lambda }-\epsilon_1 h_{\lambda
}(s)+\epsilon_2 \left(1+v_{\mu }(s)\right)\right)\right. \nonumber
\\ \left. \times \prod_{s'\in Y_{\mu}}\left( a_{\mu \lambda
}+\epsilon_1 \left(1+h_{\mu }(s')\right)-\epsilon_2 v_{\lambda
}(s')\right)\right). \label{generaldet}
\end{eqnarray}
The repercussions of this formula will be discussed in a
subsequent publication.

\section{Acknowledgements}

This work has been supported by the Volkswagen Foundation of
Germany. R. P. gratefully acknowledges the partial financial
support of INTAS 00-561 and  the Swiss grant SCOPE.


\begin{thebibliography}{99}
\bibitem{sw1}N. Seiberg, E. Witten, Nucl.Phys. B426 (1994) 19-52;
Erratum-ibid. B430 (1994) 485-486, hep-th/9407087.
\bibitem{dorey1}N. Dorey, V.V. Khoze, M.P. Mattis,
Phys.Rev. D54 (1996) 2921-2943, hep-th/9603136;
\bibitem{finnel}D. Finnell, P. Pouliot, Nucl.Phys. B453 (1995) 225-239,
hep-th/9503115;
\bibitem{ito}K. Ito, N. Sasakura, Phys.Lett. B382 (1996):95-103,
hep-th/9602073;
\bibitem{dorey3}N. Dorey, V.V. Khoze, M.P. Mattis,
Nucl.Phys. B513 (1998) 681-708, hep-th/9708036;
\bibitem{adhm}M. Atyah, V. Drinfeld, N. Hitchin, Yu. Manin,
Phys. Lett. A65 (1978) 185;
\bibitem{FPS1} R. Flume R. Poghossian H. Storch, The coefficients of
the Seiberg-Witten prepotential as intersection numbers (?),
published in the collection ``From Integrable Models to Gauge
Theories'' (World Scientific, Singapore, 02) hep-th/0110240;
\bibitem{FPS2} R. Flume, R. Poghossian, H. Storch,
Mod.Phys.Lett. A17 (2002) 327-340, hep-th/0112211;
\bibitem{H}T. Hollowood, hep-th/0201007,0202197;
\bibitem{NSch}N. Nekrasov, A.Schwarz, Commun.Math.Phys.198(1998),689 ;
\bibitem{B} J.-M. Bismut, Localization Formulas, Superconnections
and the Index Theorem of Families, Commun.Math.Phys. 103 (1986), 127-166;
\bibitem{BGV}N. Berline, E. Getzler and M. Vergne,
Heat Kernels and Dirac Operators, Springer, Berlin, 1996;
\bibitem{DK}S. K. Donaldson, P. B. Kronheimer, The Geometry of
Four-Manifolds, Oxford University Press, 1990;
\bibitem{Nak}H. Nakajima, Lectures on Hilbert schems of
Points on Surfaces, American Mathematical Society, (University Lecture Series v18) 1999;
\bibitem{N} N.Nekrasov, Seiberg-Witten Prepotential from Instanton Counting, hep-th/0206161;
\bibitem{MNSch1}G.Moore, N.Nekrasov and S.Shatashvili, Commun.Math.Phys. 209 (2000) 97-121, hep-th/9712241;
\bibitem{MNSch2}G. Moore, N. Nekrasov and S. Shatashvili, Commun.Math.Phys. 209 (2000) 77-95, hep-th/9803265;
\bibitem{DKM'}N. Dorey, T. Hollowood, V. Khoze, M. Mattis, The Calculus of Many Instantons,
hep-th/0206063;
\bibitem{HKLR}Hitchin, N.J., Karlhede, A., Lindstroem, U., Rocek, M., Hyperkaehler Metrics
and Supersymmetry, Commun. Math. Phys. 108 (1987),535-589;
\bibitem{BF} U. Bruzzo, F. Fucito, A. Tanzini, G. Travaglini, Nucl. Phys. B 611 (2001)205-226, hep-th/0008225.
\end{thebibliography}
\end{document}